\documentstyle[emulateapj,psfig]{article}

\def\be{\begin{equation}}
\def\ee{\end{equation}}
\def\ba{\begin{eqnarray}}
\def\ea{\end{eqnarray}}
\def\Mesz{M\'esz\'aros}
\def\siml{\lower4pt \hbox{$\buildrel < \over \sim$}}
\def\simg{\lower4pt \hbox{$\buildrel > \over \sim$}}
\slugcomment{Accepted for publication in The Astrophysical Journal}

\begin{document}

\title{Gamma-ray bursts with continuous energy injection and
their afterglow signature}

\author{Bing Zhang \& Peter M\'esz\'aros}
\affil{Astronomy \& Astrophysics Dept.,
 Pennsylvania State University, University Park, PA 16803}

\begin{abstract}
We consider generically the problem of a Gamma-ray burst (GRB)
fireball with an additional energy injection, either in the form
of a Poynting-flux-dominated outflow or a kinetic-energy-dominated
matter shell injected after the burst. Generally, a total injection
energy comparable to that of the impulsive energy in the initial
fireball is required to make a detectable signature in the afterglow
lightcurves. When this criterion is met in the case of
Poynting-flux-dominated injection, this leads to a gradual achromatic
bump appearing in the otherwise power-law afterglow lightcurve.
Alternatively, in the case when the injection is kinetic-energy-dominated,
the results depend on whether the collision between the rear (injected)
and the forward shell is mild or violent. If the relative velocity
between the colliding shells does not exceed a critical value defined by
their energy ratio, the collision is mild, and the injection may be analogous
to the Poynting-flux injection case. Otherwise, the injection is violent,
and an additional pair of shocks will form at the discontinuity between
two colliding shells, so that there are altogether three shock-heated
regions from which the emission contributes to the final lightcurves.
We describe the shell-merging process in detail including collision
and relaxation by taking into account the dynamical evolution and the
emission from the various shocks involved. Assuming synchrotron emission,
we calculate afterglow lightcurves in the X-ray, optical and radio bands
for the various cases. The injection signatures due to violent
matter-dominated collisions are abrupt and complicated, due to the
emission from any of the three emitting regions, and depending on the
injection parameters and the observed energy bands. This differs from
the gradual bump signature found in the Poynting-flux
injection case. In both the Poynting-flux-dominated and the
kinetic-energy-dominated cases, the energetics of the fireball as well
as the absolute afterglow flux level after the injection are boosted
with respect to the one without post-burst injection.
Identifying the different injection signatures from future early
afterglow observations may provide diagnostics about the nature of
the fireball and of the central engine.
\end{abstract}


\section{Introduction}
\label{sec:intro}

Classical Gamma-ray burst (GRB) afterglow models generally invoke an
``impulsively'' injected (possibly collimated) fireball running into
an ambient interstellar medium (ISM) or into a pre-burst stellar wind.
In principle, after the initial fireball starts to decelerate, additional
injection of energy is still possible. There are at least three motivations
to study the post-burst injection possibility more carefully.
1: Additional injection is a natural expectation of the internal shock
GRB model, which predicts that some slow shells will trail the outermost
fast shell and run into it when the latter is decelerated (e.g. Rees \&
\Mesz~ 1998; Kumar \& Piran 2000). In some cases, the central engine may
also inject fast shells at a later time. Later injection is also naturally
expected in certain types of the central engine models (e.g. Dai \& Lu
1998; Zhang \& \Mesz~ 2001a).  2: A GRB fireball can in principle be
dominated by the kinetic energy of the baryons or by a Poynting-flux
component. Any clue about the fireball nature is presently lacking. As
discussed below, an injection signature in the GRB lightcurve may directly
provide diagnostics about the nature of the injection as well as information
about the central engine. 3: ``Bump'' features have been observed in several
GRB afterglows (e.g. GRB 970228, GRB 970508, GRB 980326, GRB 000103C, etc.),
and various interpretations have been proposed, e.g., refreshed shocks
(Panaitescu, \Mesz~ \& Rees 1998), supernova components (Bloom et al. 1999;
Reichart 1999; Galama et al. 2000), dust echos (Esin \& Blandford 2000),
and gravitational micro-lensing events (Garnavich, Loeb \& Stanek
2000). It would be of great interest to study the injection feature in
more detail to find distinct properties to be differentiated from
other interpretations. Such injections are likelier to occur in the
early afterglow phase. The planned future broad-band GRB mission {\em
Swift} will have the ability of recording broadband early afterglow
signals from many GRBs, and thus bring unique opportunities to study
the injection features.

Depending on the different types of the central engine, the post-injection
could consist of, e.g., some kinetic-energy (i.e. baryon) dominated shells,
or a Poynting-flux-dominated wind (Usov 1994; \Mesz~ \& Rees 1997b).
Here we present a generic study on various injection cases. Generally
speaking, an injection energy comparable to the initial impulsive
energy is required to make the injection noticeable. We therefore
study a system containing an impulsive shell which is already heated
during the shell-ISM interaction, and which is collecting material
from the ISM, and in the meantime also receives a large enough injection
energy from a continuous Poynting wind or a trailing matter shell.
We first discuss in \S2 the simplest case, in which the fireball
continuously receives pure energy with negligible baryon loading, very
likely in the form of Poynting flux. Since no reverse shock is
expected in such a case, the injection signature is solely from the
forward shock emission, through the change of the blastwave global
dynamics. In \S3, we discuss the more complicated matter shell
injection case. In real situations, shell collisions may occur
more than once. Here we only investigate the detailed physics of one
such post-collision. The same analysis, when applied several times, is
generic in delineating the more complicated multi-collision cases.
We first analyze the general hydrodynamics of the three-shell (impulsive
shell, injective shell, and ISM) interaction (\S3.1). After reviewing
the general shell evolution history, we solve numerically for the
condition of a ``violent injection'', in which an additional pair of
strong shocks form at the discontinuity of the two colliding shells (\S3.2).
We then outline in \S3.3 the hydrodynamics of the three-shell-interaction
process in detail by dividing the whole process into five stages
(\S3.3.1-\S3.3.5), namely (1) pre-collision; (2) mild  injection;
(3) violent injection; (4) relaxation; (5) post-relaxation. In \S3.4,
we discuss synchrotron radiation from all the possible shocked regions
in various stages, and present the injection signatures in the broadband
lightcurves. Our findings are summarized and discussed in \S4.

\section{Injection from a Poynting-flux-dominated flow}
\label{sec:p}

\begin{figure*}
\centerline{\psfig{file=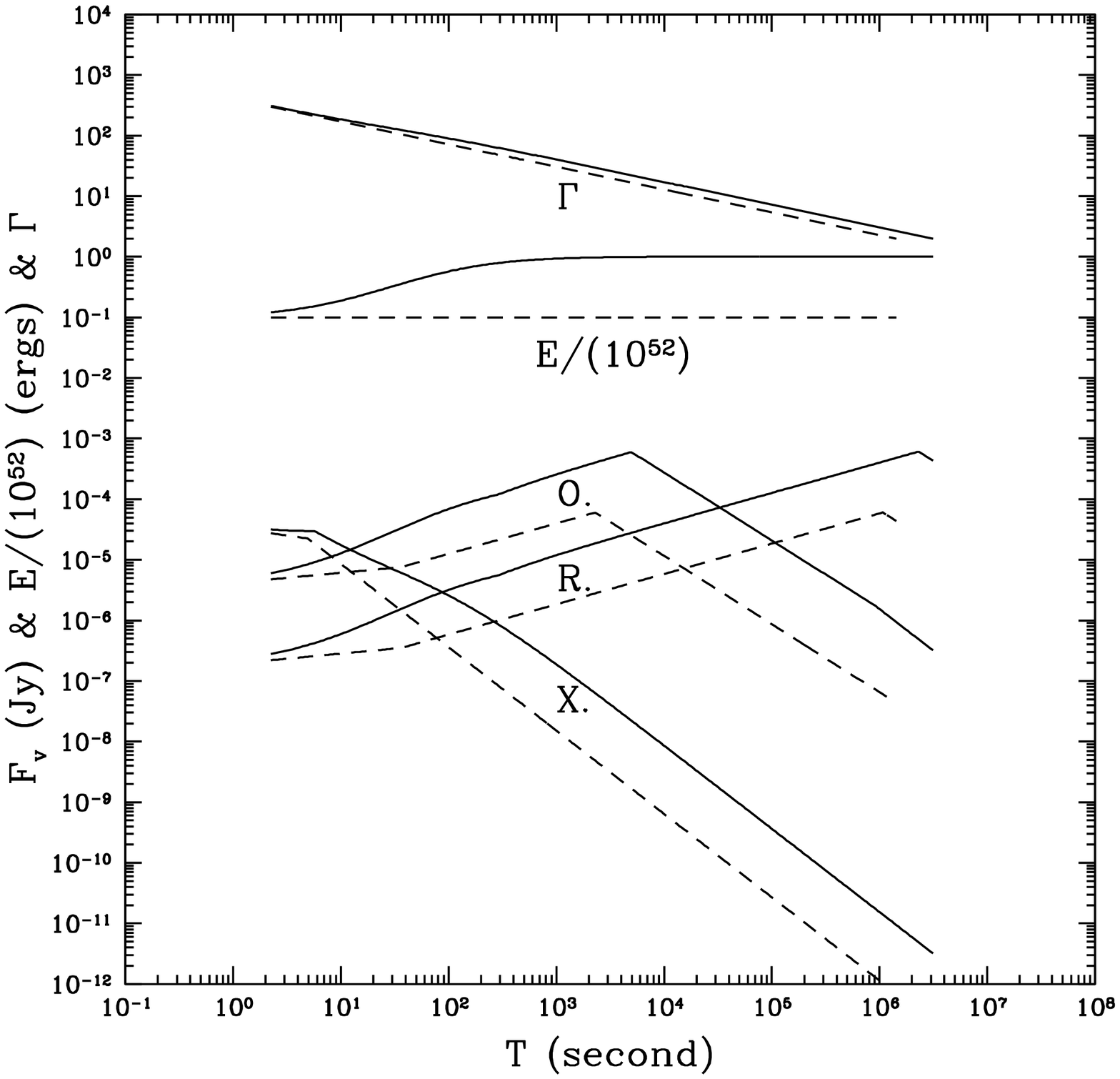,height=8.0cm}
\psfig{file=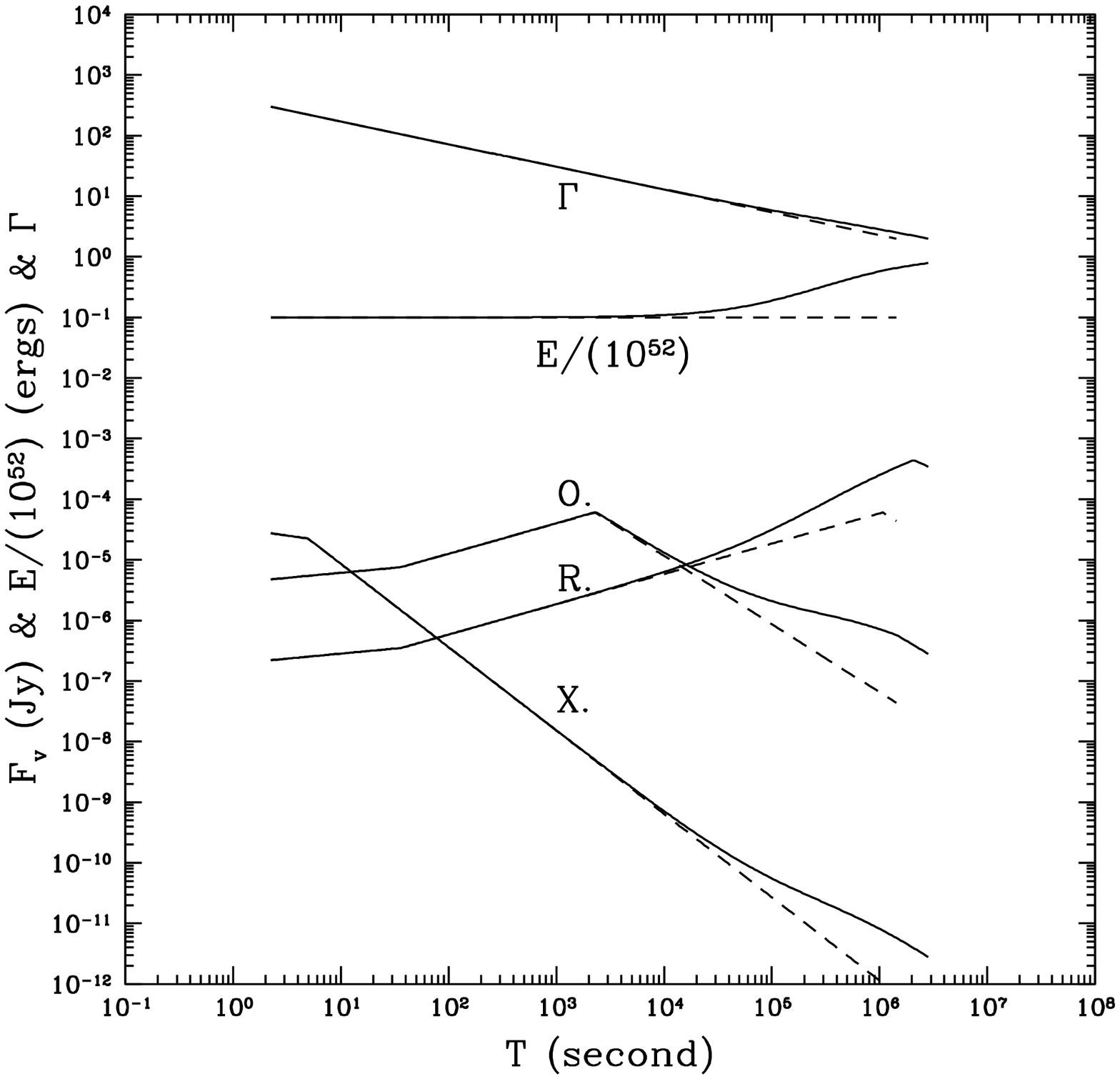,height=8.0cm}}
\caption{Injection lightcurves from a Poynting-flux-dominated flow ejected 
by a highly magnetized millisecond pulsar (magnetar) central engine.
Solid and dashed lines are the cases with or without a continuous
energy injection. Shown are the bulk Lorentz factor of the blastwave, the 
total energy in the fireball, as well as the lightcurves in various bands
(X-ray: $\nu=10^{18}$ Hz; optical: $\nu=10^{14}$ Hz; radio:
$\nu=10^{10}$ Hz). Parameters adopted:
$E_{\rm imp}=10^{51}$ ergs,  $n=1~{\rm cm^{-3}}$, $\Gamma_0=300$,
$\epsilon_e=0.1$, $\epsilon_B=0.01$, and $p=2.5$. Different pulsar
parameters lead to the injection being either ``immediate'' or
``delayed''. (a) Immediate injection: $B_p=10^{16}$ G and $P_0=1.5$
ms; (b) delayed injection: $B_p=10^{14}$G and $P_0=1.5$ ms.}
\end{figure*}

The conventional definition of a Poynting-flux-dominated flow is
$L_{\pm,\gamma,N} /L_P \ll 1$, where $L_{\pm,\gamma,N}$ is the luminosity
of the electron-position pair, radiation and nucleon components, and $L_P$
is the luminosity of the Poynting-flux (i.e. magnetic field) dominated
outflow component (e.g. Usov 1999).
In the context of GRBs, discussions of a Poynting-dominated flow as
compared to a kinetic-energy-dominated flow are mainly focused on
the paradigm of an approximately baryon-free Poynting flow, which are
not as conducive to collisionless shock formation, so in this case GRBs
may be powered by the non-linear breakdown of large-amplitude electromagnetic
waves (Usov 1994; Blackman \& Yi 1998; Lyutikov \& Blackman 2001; for a
review, see Usov 1999). However, in both scenarios, collisionless external
forward shocks are believed to form in the afterglow phase (Usov 1999).
Although the baryon free nature at the GRB prompt phase is uncertain (e.g.
Spruit, Daigne \& Drenkhahn 2001), in the later injection phase the bayron
loading could be in principle much lower and the injection could be in the
form of an almost bayron-free Poynting-flux-dominated wind. In such a
case, no collisionless reverse shock is expected to propagate into the
wind (Kennel \& Coroniti 1984, Usov 1999; essentially, because the
sound speed in the ejecta is close to the speed of light), and the
injection energy is used to increase the total energetics of the fireball.

Considering an adiabatic relativistic hot shell which is collecting
material as it expands into the ISM and which receives an energy
input from the central engine via a Poynting-flux-dominated flow,
the differential energy conservation equation in a
fixed frame is $d[\Gamma(M_0 c^2+M_2c^2+U)]=d M_2c^2
+ d E_{\rm inj}$, where $M_0$, and $M_2$ are the masses of the
impulsive shell and the swept-up ISM, respectively\footnote{The
subscript 2 is used to match the notation in the
kinetic-energy-dominated case discussed in \S3, e.g., Figure 2 and
equation (\ref{stg:2}).}, $E_{\rm inj}$ is the received injection energy,
and $\Gamma$ is the bulk Lorentz factor of the blastwave. The injection
wind can be approximated as being cold, so it does not provide any
internal energy, and the total internal energy in the fireball is provided
by the shock heated ISM.
Noticing equation (\ref{21}), one can usually
approximate $U\simeq (\Gamma-1)M_2c^2$ as a working assumption to
perform a first-order treatment of the problem. Strictly speaking,
such an approximation is only valid when the post-shock fluids are
considered uniform, i.e., with constant density, temperature, and
Lorentz factor. More detailed self-consistent modelings require taking
into account the Blandford-McKee (1976) self-similar profiles and
tracking the post-shock internal energy more closely. With the assumption,
the integrated energy conservation equation then reads
\begin{equation}
E_{\rm inj}+(\Gamma_0-\Gamma)M_0c^2=(\Gamma^2-1)M_2c^2.
\label{E}
\end{equation}
This is analogous to the combination of equations (1) and (2) in Zhang
\& \Mesz~ (2001a). When $E_{\rm inj}$ is negligible and noticing that
$E_{\rm imp}\simeq \Gamma_0 M_0c^2$, equation (\ref{E}) is the
standard equation for the impulsive adiabatic blastwave evolution
(e.g. Chiang \& Dermer 1999; Huang, Dai \& Lu 1999), which can be
reduced to the familiar form $E_{\rm imp}=(4\pi/3)R^3nm_pc^2
\Gamma^2$ (so that $\Gamma \propto R^{-3/2} \propto T^{-3/8}$, where
$R$ is the blastwave radius in the fixed frame, and $T$ is the observer
time) when $\Gamma_0 \gg \Gamma \gg 1$. In the presence of $E_{\rm inj}$,
we can see that the condition for changing the blastwave dynamics is that
$E_{\rm inj}$ exceed $(\Gamma_0-\Gamma)M_0c^2$, or essentially
$E_{\rm inj} \simg E_{\rm imp}$.

A Poynting-flux-dominated wind is usually continuous, with an intrinsic
luminosity law which may be defined through a temporal index $q$ within
a certain time regime, e.g. $L(T) \propto T^q$, where $T$ is the
intrinsic time of the central engine (which is the same as the
observer's time after the cosmological time dilation correction).
A detailed description of such an injection scenario, especially in the
context of magnetar central engine models, was discussed in a previous
paper (Zhang \& \Mesz~ 2001a). Here we summarize some of the conclusions.
When the blastwave dynamics is dominated by the injection (i.e. $E_{\rm inj}
\simg E_{\rm imp}$, which only occurs when $q>-1$), the bulk Lorentz
factor evolves according to $\Gamma \propto R^{-(2-q)/2(2+q)}\propto
T^{-(2-q)/8}$. In most problems, usually $q=0$ during the injection
phase, so that $\Gamma \propto R^{-1/2}\propto T^{-1/4}$.
There are three relevant time scales in the problem, i.e., the blastwave
deceleration time, $T_0$; the critical time when the continuous-injection
energy component dominates the impulsive-injection energy component, $T_c$;
and the characteristic time durung which the central engine produces a
substantial energy imput (i.e. $q>-1$), ${\cal T}$.
The condition for the injection signature to show up in the
afterglow lightcurves is ${\cal T}> {\rm max} (T_c, T_0)$. For
$T_c < T_0 < {\cal T}$, the dynamics is defined by the
continuous-injection as soon as the afterglow is set up, and we
define it as an immediate injection (Fig.1a). The
lightcurves will be flat from $T_0$ to ${\cal T}$, and will steepen
after ${\cal T}$. For $T_0 < T_c < {\cal T}$, the signature will not
show up until $T_c$, and we define it as a delayed
injection (Fig.1b). There are two temporal index changes: the
lightcurves will flatten around $T_c$, and resume the original
steepness around ${\cal T}$. The injection-signature is achromatic due
to the change of the blastwave dynamics. Since no reverse shock is
expected, only emission from the forward shock is responsible for the
afterglow emission. Assuming synchrotron emission of the relativistic
electrons, the temporal index at the injection phase is $\alpha =
(1-q/2)\beta+ q+1$ (which is $\beta+1$ for $q=0$), as compared to
the standard case of $\alpha=(3/2) \beta$, where $\beta$ is the spectral
index in a certain band of interest.
As an example, we again discuss the Poynting-flux injection
case in the magnetar model. In this particular case, the
injection luminosity law is
$L(T)=L_{em,0} (1+T/{\cal T}_{em})^{-2}$, where
$L_{em,0}={I\Omega_0^2}/{2{\cal T}_{em}}\simeq 1.0\times10^{49}
{\rm erg~s^{-1}}B_{p,15}^2P_{0,-3}^{-4}R_6^6$ and ${\cal
T}_{em}={3c^3I}/{B_p^2R^6\Omega_0^2} =2.05\times10^3~{\rm s}~
I_{45}B_{p,15}^{-2}P_{0,-3}^2R_6^{-6}$, and $B_p=10^{15} {\rm cm}
B_{p,15}$, $P_0 =10^{-3} {\rm s} P_{0,-3}$ are the pulsar dipolar
field strength at the pole, and the initial period at birth,
$I=10^{45} {\rm g~cm^2} I_{45}$ and $R=10^6 {\rm cm} R_6$ are the
rotation inertia and the radius of the pulsar, respectively.
Figure 1 shows two example injection lightcurves for both the
immediate and the delayed injection cases in three energy bands.
A general feature is that the injection signature is rather smooth,
mainly due to the smooth varying luminosity law $L(T)$. Dai \& Lu (2001)
and Wang \& Dai (2001) obtained a similar result recently.

\section{Injection from a kinetic-energy-dominated shell}
\label{sec:k}

The previous Poynting-dominated injection is, for the present purpose, the
simplest case, in the sense that the injection does not result in new
emission sites. The signature in the lightcurves is still produced by
the shocked ISM, albeit through a changed global dynamics of the blastwave.
However, if the injection energy is dominated by the kinetic energy of the
baryons, collisionless shocks may in principle form at the discontinuity of
any two adjacent shells under certain conditions. The emission sites during
the injection process therefore could be multiple, and the injection
signatures, due to the joint contributions from all these sites, should be
much more complicated. In this section, we investigate in some detail
the shell collision process, as well as the emission signatures.

\begin{figure*}
\centerline{\psfig{file=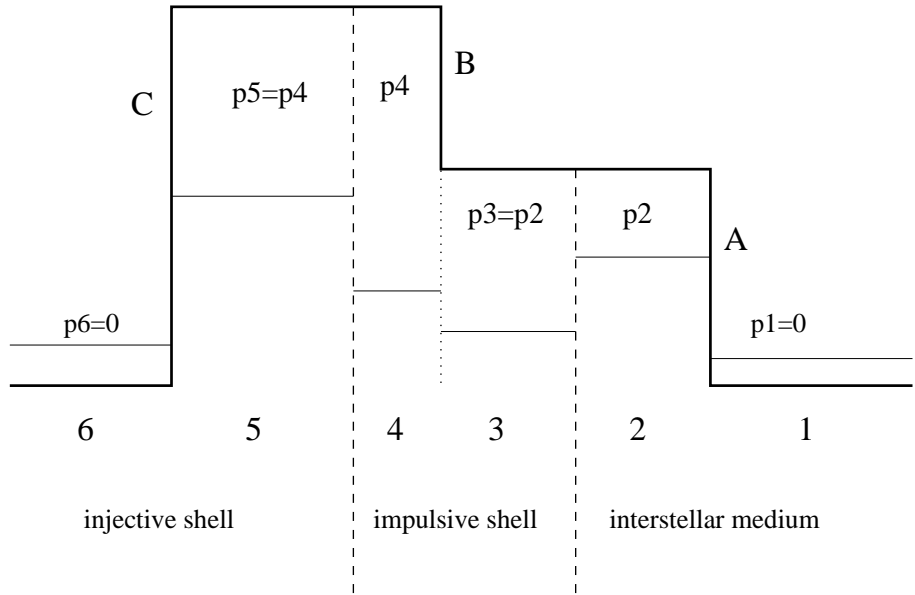,angle=-90,height=8.0cm}}
\caption{A schematic (not to scale) description of the violent injection
configuration from a kinetic-energy-dominated shell. The thick solid
line indicates the pressure, or the internal energy density, in each
region. The thin solid line indicates the co-moving mass density in
each region.}
\end{figure*}

\subsection{Hydrodynamics}

We consider a collision in the afterglow phase, in which case the
impulsive shell is already heated up by the passage of the
reverse shock. Assuming that the electron equipartition factor
$\epsilon_e$ is not close to unity, the impulsive shell will
not cool at the time of injection, even if the electrons cool
rapidly. A generic configuration of the problem is summarized in Figure 2,
which is predicated upon the fulfillment of certain conditions (\S3.2).
There are six regions of interest, namely: (1) unshocked ISM; (2)
forward-shocked ISM; (3) unshocked hot impulsive (leading) shell; (4)
forward-shocked impulsive shell; (5) reverse-shocked injective
(trailing) shell; (6) unshocked injective shell.  We denote through
$M_i$, $n_i$, $e_i$ and $p_i$ the mass, baryon number density, internal
energy density, and the pressure of the region ``$i$'' in its own
rest frame, respectively; $\gamma_i$ and $\beta_i$ are the bulk
Lorentz factor and the dimensionless velocity of region ``$i$''
measured in the fixed frame, respectively; and
\be
\gamma_{ij} \simeq (\gamma_i/\gamma_j+\gamma_j/\gamma_i)/2,
~~{\rm for}~~ \gamma_{i,j} \gg 1
\label{gamij}
\ee
and $\beta_{ij}$ the bulk Lorentz factor and the dimensionless
velocity of region ``$i$'' measured in the rest frame of region ``$j$'',
respectively. In the hot regions (2, 3, 4 and 5), we assume a
relativistic equation-of-state, i.e., $p_i= (\hat \gamma-1)e_i=e_i
/3$, where $\hat \gamma=4/3$. Two contact
discontinuities separate the three parts, i.e., the ISM, the
impulsive shell (which we will consistently call shell 3 in the
following discussions), and the injective shell (which we call shell 6
in the following discussions). The pressure equilibrium and
velocity continuity at the contact discontinuities yield
\ba
e_3=e_2,  & \gamma_3=\gamma_2,  \label{e32}\\
e_5=e_4,  & \gamma_5=\gamma_4.  \label{e54}
\ea
There are three shocks that accelerate particles and contribute to the
emission, i.e., (A) a first forward shock propagating into the ISM,
(B) a second forward shock propagating into the impulsive shell, and
(C) a reverse shock propagating into the injective shell.
The continuity of the energy, momentum, and particle flux densities in
the rest frame of each shock leads to the following jump conditions
(Blandford \& McKee 1976; Sari \& Piran 1995; Kumar \& Piran 2000)
\be
\frac{e_2}{n_2}=(\gamma_2-1) m_pc^2, ~~~ \frac{n_2}{n_1}=4\gamma_2+3,
	\label{21}
\ee
\be
\gamma_{43}^2=\frac{(1+3e_4/e_3)(3+e_4/e_3)}{16 e_4/e_3},  \label{gam43}
\ee
\be
\left({n_4 \over n_3}\right)^2=\frac{(e_4/e_3)(1+3e_4/e_3)}
	{3+e_4/e_3},  \label{n43}
\ee
\be
\frac{e_5}{n_5}=(\gamma_{56}-1) m_pc^2, ~~~
	\frac{n_5}{n_6}=4\gamma_{56}+3~.
\label{56}
\ee
Notice that the jump conditions
at the second forward shock (eqs. [\ref{gam43}] and [\ref{n43}]) are
different from the others, mainly because the region in front of the shock
(region 3) is hot. With (\ref{e32}) and
(\ref{e54}), a combination of (\ref{21}) and (\ref{56}) yields
\be
\frac{e_4}{e_3}=\frac{n_6}{n_1} \frac{F(\gamma_{56})}{F(\gamma_3)}
\simeq \frac{E_{0,6}}{E_{0,3}}\cdot \frac{F(\gamma_{56})}{4}
\cdot {\rm min}\left(1,\frac{R}{R_{s,6}}\right),
\label{e4/e3}
\ee
where
\be
F(\gamma)=(\gamma-1)(4\gamma+3) \simeq
4\gamma^2, ~~~ \gamma \gg 1~.
\label{F}
\ee
The second half of the equation (\ref{e4/e3}) will be explained further
below. Given $\gamma_3$ and $\gamma_6$, the unknown $\gamma_4$ can be
solved for by replacing (\ref{e4/e3}) into (\ref{gam43}) with the use of
(\ref{gamij}).

\subsection{Shell evolution \& condition for violent collision}

Equations (\ref{21})-(\ref{56}) imply that
strong shocks form between the contact discontinuity of the injective
shell and the impulsive shell (e.g. Blandford \& McKee 1976), as
indicated in Figure 2. A physical solution for the non-linear
equations (\ref{21})-(\ref{56}) requires $e_4 > e_3$, or $\gamma_3
< \gamma_4 < \gamma_6$, which demands a minimum relative Lorentz
factor between the two colliding shells given the parameters of both
shells (eq.[\ref{e4/e3}]). This is qualitatively in agreement with 
Kumar \& Piran (2000), who demand a minimum enthalpy density ratio 
for a given $\gamma_{36}$ (in their Fig.3), which is equivalent to 
our case in which we demand a minimum
$\gamma_{36}$ for a given $E_{0,6}/E_{0,3}$.
For a shock to form, the injective
shell should move super-sonically with respect to the
impulsive shell, i.e., by noticing the sound speed $c_{s} \simeq
c/\sqrt{3}$ for the shell 3 (in the relativistic limit $\hat
\gamma=4/3$), $\gamma_{36} \geq \sqrt{3/2} \simeq 1.22$. Usually this
condition is automatically satisfied given the strong shock forming
condition $e_4>e_3$.
In this subsection, we examine the
criterion for $e_4>e_3$, which is the condition for the configuration
in Figure 2 being realized. We first summarize the evolution of a matter
fireball shell (e.g. \Mesz, Laguna \& Rees 1993; Piran, Shemi \&
Narayan 1993; Kobayashi, Piran \& Sari 1999; Piran 1999), and then
show how a shell collision process is well-defined by the initial
conditions of the shells.

\begin{table*}
\centerline{\psfig{file=table1.epsi}}
\end{table*}

The injection of a matter shell from the central engine
can be quantified by three independent parameters, i.e., the energy of
the shell $E_{0}$, the mass of the shell $M_{0}$ (thus the dimensionless
entropy, or the maximum Lorentz factor of the shell is
$\eta=\Gamma_0 = E_{0} / M_{0}c^2$), and the initial thickness of the
shell $\Delta_{0}$ which is essentially determined by the timescale of
the central engine activity, i.e., $\Delta_0 = c \delta T$.
The following parameters can be consequently defined:  the initial baryon
number density of the shell, which is $n_0=M_{0} /[m_p (4\pi/3)\Delta_0^3]$;
the radius at which the fireball stops accelerating and enters the coasting
phase, which is $R_c \simeq \Gamma_0 \Delta_0$; and the radius at which
the fireball starts to spread with its local sound speed
due to the velocity difference within the fireball matters, which is
$R_s \simeq \Gamma_0^2 \Delta_0$. There is another parameter, i.e.,
the deceleration radius $R_d$, which usually appears in the problem.
However, this radius not only depends on the intrinsic parameters of
the shell (e.g. total energy), but also on the properties (e.g. density
and velocity) of the medium that stops the shell. For the interaction
between the impulsive shell (shell 3) and the ISM, the deceleration radius
is roughly the radius where the mass of the ISM collected by the fireball
is $1/\gamma_{0,3}$ of the shell mass, i.e.,
$R_{d,3} \sim R_{\gamma,3} = (3E_{0,3}/4\pi n_1 m_p c^2
\eta_{3}^2)^{1/3}=1.2 \times 10^{16} {\rm cm}~ E_{0,3,51}^{1/3}
n_1^{-1/3} (\gamma_{0,3}/300)^{-2/3}$, where $n_1$ is the ISM baryon
number density in unit of 1${\rm cm}^{-3}$, $E_{0,3,51}=E_{0,3}/10^{51}
{\rm ergs}$ (hereafter $X_n=X/10^n$, except those subscripts that denote
different regions, such as $n_1$).
For the interaction between the injective shell (shell 6) and the
impulsive shell (shell 3), the deceleration
radius is around the radius where the injective shell catches
up with the impulsive shell. Let us denote $\Delta T_{36}$
as the time interval between ejecting the two adjacent shells (as
compared to $\delta T$, which is the timescale of the injection
process), the ``catching-up'' radius is then $R_{d,6} \simeq 2
\gamma_3^2 c (\Delta T_{36}) / (1-\gamma_3^2/\gamma_6^2)$. For
$\gamma_6 \gg \gamma_3$, $R_{d,6}$ is mainly defined by the Lorentz
factor of the slow, leading shell, i.e., $R_{d,6} \simeq
2\gamma_3 ^2 c (\Delta
T_{36})=2.1\times 10^{16} {\rm cm}~E_{0,3,51}^{1/4} (\Delta T_{36} /
20 {\rm s})^{1/4} n_1^{-1/4}$. The condition for the
collision to occur in the afterglow phase is then $R_{d,6} > R_{d,3}$,
or approximately $\Delta T_{36} > 2.3 {\rm s}~ E_{0,3,51}^{1/3}
n_1^{-1/3} (\gamma_{0,3}/300)^{-8/3}$. The deceleration radius of a shell
can in principle be either smaller or larger than the spreading radius,
$R_s$. For $R_s > R_d$ (a ``thick shell'', e.g. Sari \& Piran 1995),
the shell has not started to spread when the deceleration (or the
collision) occurs. Alternatively, for a ``thin shell'' (with $R_s <
R_d$), the shell is spreading with its sound speed in the comoving
frame due to the velocity difference within the shell as the collision
occurs. In this sense, the initial ``thin shell'' is actually thick during
the collision. Table 1 summarizes the evolution of the shell widths
(both in the fixed frame, $\Delta$, and in the comoving frame, $\Delta'$),
shell baryon number density $n$, and the ratio $n/n_1$ in various regimes.
In the last column
(i.e., $n/n_1$), for the rear shell 6, we have assumed that the radius
of the shell 6 front is comparable to that of the shell 3 front, i.e.,
we assume that $\Delta_3 \ll R$, which is generally the case as long
as $\gamma_3 \gg 1$. We note that expressing $n/n_1$ in terms of
the mass (or energy) ratio of the shell and the collected ISM ($M_2$)
allows a clearer physical understanding for the problem.
At the collsion point $R_{d,6}$, it is convenient to show
that $n_6/n_1 =(M_{0,6}/M_2) \cdot {\rm min}(R/R_{c,6},
\gamma_{0,6})$ (thick shells for the former, and thin shells for the
latter). Noticing that $\gamma_{6} M_{6,0} = E_{\rm inj}=E_{0,6}$, and
that for $\gamma_{0,3}\gg \gamma_3 \gg 1$ (i.e., the collision occurs
long after the afterglow's setting-up but well before the
non-relativistic phase, which is usually the case), one has
$F(\gamma_3) \simeq 4 \gamma_3^2$, and $\gamma_3^2 M_2 \simeq E_{\rm
imp}=E_{0,3}$ (see eq.[\ref{stg:2}] and the discussions
thereafter). This justifies the second part of the equation
(\ref{e4/e3}).

The condition for $e_4>e_3$ can be investigated numerically. Figure 3
shows the minimum $\gamma_{36}$ required for $e_4>e_3$ for different
energy ratios of the two shells, $E_{0,6}/E_{0,3}$. Notice that the
wiggles on the curves are due to the crudeness of the numerical
solutions. A general trend is that a larger relative velocity between
the two shells is required for a less energetic trailing shell. The
critical line depends on the status of the shell 6, i.e., whether it
is expanding (thin shell) or not (thick shell) when the collision occurs.
This is least demanding for an expanding (thin) shell 6, which is plotted
as the solid curve. For a thick shell 6, the larger the ratio $R_s/R$,
the more demanding the condition is. This is because for the same energy
ratio $E_{0,6}/E_{0,3}$, a thicker shell tends to get a smaller
$e_4/e_3$ ratio (eq.[\ref{e4/e3}]). The dotted and the dashed curves
in Figure 3 are for a shell with $R_{s,6}$ 5 and 10 times the
collision radius, $R_{d,6}$, respectively. Another comment is that,
as long as in the relativistic phase lasts, only the energy ratio and
the relative velocity are relevant to the problem. The absolute values
of the shell energies and Lorentz factors do not enter the solution.
Notice that the supersonic condition defines another
critical line for the shock forming condition (Fig.3).
The case for $e_4 >e_3$ not being satisfied is less clear, but in this
case no strong shock is expected. We regard such
an injection case as ``mild'' (with weak or non-shock), and we expect
that the injective materials are effectively attached onto the
forward shell to increase the total energy budget, which is analogous
to the Poynting-flux injection case.

In summary, the input parameters in the injection problem include:
(i) $E_{\rm imp} =
E_{0,3}=\gamma_{0,3} M_{0,3}c^2$; (ii) $\gamma_{0,3}$ (or $M_{0,3}$,
which is only important around the on-set of the afterglow); (iii)
$n_1$; (iv) $E_{\rm inj}=E_{0,6}= \gamma_{0,6}M_{0,6} c^2$; (v)
$\gamma_{0,6}$ (or $M_{0,6}$); (vi) $\Delta_{0,6} = c \delta T_6$,
which determines $R_{s,6}=\gamma_{0,6}^2 \Delta_{0,6}$; (vii)
$\Delta T_{36}$ (or equivalently $R_{d,6}$). The comparison between
$R_{s,6}$ and $R_{d,6}$ then decides whether the trailing shell is
``thin'' or ``thick'' when the collision occurs. Given a set of
injection parameters (i) - (vii), at any radius $R$, one can first
check the shock forming condition in Figure 3. If the condition is not
satisfied (below the critical line), the injection is a ``mild'' one.
If the condition is satisfied (above the critical line in Fig.3), the
shell-merging process is violent.

\begin{figure*}
\centerline{\psfig{file=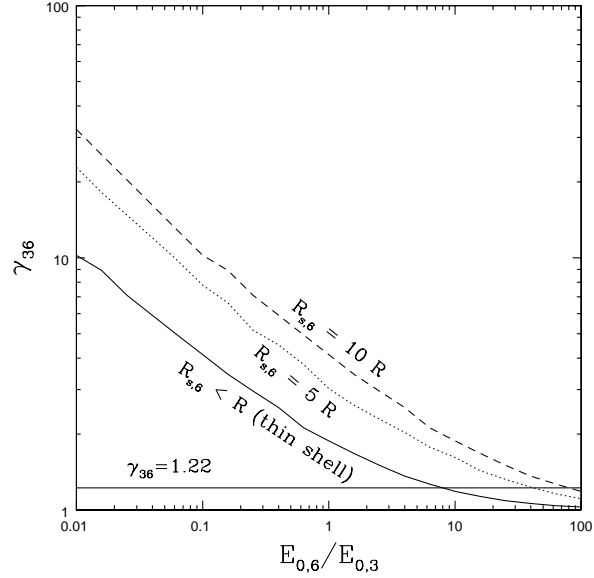,height=8.0cm}}
\caption{The condition for a violent injection. The
three curves are the critical lines for $e_4>e_3$ being satisfied, for
different cases of the rear shell (thin or thick at the collision).
The super-sonic condition (the horizontal line) is also plotted.
The space region above both the critical line and the horizontal line
is where a violent collision as indicated in Fig.2 occurs.}
\end{figure*}

\subsection{Shell merging process}

Though an injection of matter shell could be mild or violent
throughout, in some cases the injection could be mild in the
beginning, and turn violent at larger radii when $\gamma_3$ drops so
that $\gamma_{36}$ meets the critical value.
In this subsection, we present an example which includes both cases
during the injection. There are four characteristic radii in the
problem: (1) the collision radius, $R_{col} = R_{d,6}$; (2) the radius
where the persistent shocks form (Fig.3), $R_{sh}$; (3)
the radius where the injection process ceases, $R_\Delta$; (For a mild
injection, this is the radius where all the material in the shell 6
catches up with and ``attaches'' onto shell 3; for a violent
injection, this is where the reverse shock crosses the rear shell and
the merging process enters the relaxation stage.) (4) the radius where
$e_4/e_3$ drops to unity and the relaxation finishes, $R_f$. The whole
shell-merging process may be then be (at most) divided into five stages
(Fig.4a): (1) pre-collision, $R<R_{col}$; (2) mild injection, $R_{col}
< R <R_{sh}$; (3) violent injection, $R_{sh}<R<R_\Delta$; (4) relaxation,
$R_\Delta<R<R_f$; and (5) post-relaxation, $R>R_f$. Below we discuss each
stage in some detail.

\subsubsection{Pre-collision stage}

Let us consider a central engine that ejects an impulsive
fireball\footnote{The impulsive fireball could be in principle highly erratic,
as indicated by spiky GRB lightcurves. In the afterglow phase, many
mini-shells collide and merge into one big shell, and we regard this as
equivalent to the central engine effectively ejecting one single shell.}
(shell 3) with $E_{\rm imp} = E_{0,3}=\gamma_{0,3} M_{0,3}c^2$. Some time
($\Delta T_{36}$) later, it ejects another (effective) shell (shell 6) with
$E_{\rm inj}= E_{0,6} =\gamma_{0,6} M_{0,6}c^2$. The forward shell starts
to decelerate at $R_{d,3}$ due to the interaction with the ISM, and
at a certain time before collision, it has collected a mass $M_2$ worth of
ISM, and has a Lorentz factor of $\gamma_3$. The rear shell moves faster,
and we assume that before collision it is essentially undecelerated, and is
still propagating with $\gamma_{0,6}$. The whole system has two independent
sub-systems. The deceleration of the forward shell is quantified by (cf.
eq.[\ref{E}] but without the injection term)
\be
(\gamma_{0,3} -\gamma_3)M_{0,3}=(\gamma_3^2-1)M_2~,
\label{stg:1}
\ee
which is the standard adiabatic afterglow evolution. The rear shell is
not noticeable at this stage.

\subsubsection{Mild injection stage}

Starting from $R_{col}=R_{d,6}\simeq 2\gamma_3^2 (c \Delta T_{36})$, the
shell 6 catches up with the shell 3. At $R_{col} < R <R_{sh}$, the
injection is not energetic enough to excite shocks. The injection is
analogous to the case of a Poynting-flux-dominated flow, i.e., mainly
it increases the blastwave energetics. Assuming that at a certain radius
(time) a mass of $M_5$ from the total injection shell $M_{6,0}$
reaches the shell 3 and its energy is added to the total blastwave
energy (here the subscript 5 is only for easy comparison with the
following notations, eqs.[\ref{stg:3}], [\ref{stg:4}] and
[\ref{stg:5}], but does not have
the physical meaning illustrated in Fig.2), one can write down the
global energy conservation of the system. The total energy of the two
shells and the (collected) ISM if there were no interaction among them
is $(\gamma_{0,6} M_{0,6}+\gamma_{0,3} M_{0,3}+M_2)c^2$; that with the
interactions is $\gamma_{0,6} (M_{0,6}-M_5)c^2+\gamma_3 (M_5c^2+
M_{0,3}c^2+U_{2+3+5})$. These two should be equal. Noticing that
$U_{2+3+5} \simeq (\gamma_3 -1)M_2c^2$, one has
\be
(\gamma_{0,6}-\gamma_3)M_5+M_{0,3}(\gamma_{0,3}-\gamma_3)=(\gamma_3^2
-1)M_2~.
\label{stg:2}
\ee
This is very analogous to (\ref{E}). For $\gamma_{6,0} \gg \gamma_3$,
the first term on l.h.s is approximately $\gamma_{6,0}M_5$, which is
just the energy injected into the system.

During this mild injection phase, the mass in the shell 3 gradually
increases while that in the shell 6 (un-collided part) gradually
decreases. This ends when the $e_4>e_3$ condition is satisfied at
$R_{sh}$ and a pair of shocks (B and C) propagate into the two
colliding shells, or, in a less energetic
situation, when the whole shell 6 is ``attached'' onto the shell 3 and
the injection process ceases. For the former case, we define $E_{sh,3}
= E_{0,3}+M_5c^2$ and $E_{sh,6}=E_{0,6}-M_5c^2$, both of which will be
used later in the stage 4 (\S3.3.4).

\subsubsection{Violent injection stage}

If the shell 6 is energetic enough or fast enough (see Fig.3), one then has
the configuration shown in Figure 2. The two new shocks (B and C)
propagate into the two colliding shells, respectively. In principle,
one can define this stage to end when either of the two shocks crosses
the corresponding shell. However, the reverse shock always crosses the
rear shell first, and we define the radius where this happens as
$R_\Delta$. The violent injection stage is then defined as $R_{sh} < R
< R_\Delta$.

To describe the physical process at this stage, we need to quantify
the dynamical evolution of the three shocks, or equivalently, the
temporal evolution of $\gamma_3=\gamma_2$ and $\gamma_5=\gamma_4$.
Strictly speaking, $\gamma_3$ and $\gamma_4$
should be solved self-consistently, and one needs one more independent
relation between $\gamma_3$ and $\gamma_4$ besides (\ref{gam43}) and
(\ref{e4/e3}), in order to get a self-consistent solution.
However, this is usually not necessary at this stage. Again, we write
down the global energy conservation of the whole system, i.e.,
$\gamma_{0,6} M_{0,6}c^2+\gamma_{0,3} M_{0,3}c^2+M_2 c^2=
\gamma_{0,6} (M_{0,6}-M_5)c^2+\gamma_4 (M_5c^2+M_4c^2+U_{4+5})
+\gamma_3 (M_2c^2+ M_{0,3}c^2-M_4c^2+U_{2+3})$. Noticing $U_{2+3}\simeq
(\gamma_3-1)M_2c^2$, $U_{4+5}\simeq (\gamma_{56}-1)M_5c^2$, this
gives a time-dependent equation
\ba
(\gamma_{0,6}-\gamma_4\gamma_{56})M_5-(\gamma_4-\gamma_3)M_4 \nonumber\\
+(\gamma_{0,3} -\gamma_3)M_{0,3}=(\gamma_3^2-1)M_2~.
\label{stg:3}
\ea
Comparing to equations (\ref{stg:2}) and (\ref{E}), we find that the
first term in those equations (the already injected energy) is now
replaced by the first two terms in (\ref{stg:3}). The global dynamics
($\gamma_3$ evolution) is modified only when the sum of the first two
terms is comparable to the third term which is essentially
$E_{\rm imp}/c^2=\gamma_{0,3} M_{0,3}$. This is impossible, since
the energy already injected into the system (i.e. $\gamma_{0,6} M_5$)
is now partially consumed in maintaining the kinetic energy of
the injected material [i.e., $-\gamma_4\gamma_{56} M_5 c^2$, since
for $\gamma_{0,6} \simg \gamma_3$ the term $(\gamma_{0,6}-\gamma_4
\gamma_{56})$ is nearly zero], and in raising the Lorentz factor of
the newly shocked regions [i.e. the term $-(\gamma_4-\gamma_3)M_4$].
Unless $E_{\rm inj} \gg E_{\rm imp}$, which is rather unlikely in
GRBs, the sum of the first two terms is negligibly small, and
(\ref{stg:3}) is essentially reduced to (\ref{stg:1}). The simple
scaling law of $\gamma_3$ for the impulsive injection case is then
applicable at this stage, e.g., for an isotropic adiabatic fireball
running into a uniform ISM, one has $\gamma_3 \propto R^{-3/2} \propto
T^{-8/3}$.

The unknown quantity $\gamma_4$, of interest here, can then be solved for
by using (\ref{gam43}) and (\ref{e4/e3}). The temporal evolution of both
$\gamma_{34}$ and $\gamma_{56}$, as well as the random Lorentz factor
in the region 4, $\gamma_{_{\rm B}}$, are thus well quantified.
Although generally this is done numerically, a rough analytic estimate
is possible. When the solution of $\gamma_4$ is physical, one can
approximate $\gamma_6 \gg \gamma_4 \gg \gamma_3$, as long as
$\gamma_6 / \gamma_3 \gg 1$. In this limit, $\gamma_{56} \sim
\gamma_6 / 2\gamma_4$, $\gamma_{43} \sim \gamma_4/2\gamma_3$, and one
has $\gamma_4 \propto (n_6/n_1)^{1/4} \gamma_6^{1/2}$. For our assumption
of $\gamma_6 \sim \gamma_{6,0}=$const, we have $\gamma_4 \propto n_6^{1/4}
\propto R^{-3/4}$ for a spreading shell ($\propto R^{-1/2}$ for a
non-spreading shell\footnote{Hereafter the scaling for an spreading
shell is not bracketed, while that for a non-spreading shell is.}).
We can further estimate $e_4/e_3 \propto \gamma_{43}^2 \propto n_6^{1/2}
R^3 \propto R^{3/2}$ (or $\propto R^2$). Since $e_3 \propto \gamma_3^2
\propto R^{-3}$, we have $e_4 \propto n_6^{1/2} \propto R^{-3/2}$ (or
$\propto R^{-1}$). With (\ref{n43}), we also have $n_4/n_3 \propto
(e_4/e_3)^{1/2} \propto R^{3/4}$ (or $\propto R$). Noticing that the
shell 3 is spreading and that $n_3 \propto R^{-3}\gamma_3 \propto
R^{-9/2}$, one then has $n_4 \propto R^{-15/4}$ (or $\propto R^{-7/2}$).
The ``random'' Lorentz factor in the region 4 (at the shock B) which
defines the radiation from the region (notice that it is different from
$\gamma_{43}$) is then $\gamma_{_{\rm B}} = e_4/n_4m_pc^2 \propto n_6^{1/4}
R^3 \propto R^{9/4}$ (or $\propto R^{5/2}$). Finally, the random Lorentz
factor at the shock C is essentially $\gamma_{_{\rm C}}=\gamma_{56}$,
which scales as $\propto n_6^{-1/4} \propto R^{3/4}$ (or $\propto
R^{1/2}$). These very crude estimates generally agree with the numerical
results (see Fig.4a).

Finally we estimate $R_\Delta$. A self-consistent calculation of the
shock crossing time (for both shells) requires hydrodynamically
tracking the speeds of the shocked and un-shocked shells. Very
roughly, the shock crossing time (e.g., for the rear shell 6) may be
estimated (Sari \& Piran 1995) $t_{\Delta_6}=[\bar
\Delta_{6}/{c(\beta_6-\bar \beta_5)}] \cdot (1-{\gamma_{6} \bar n_6}/
{\bar\gamma_5 \bar n_5}) \sim (\Delta_6/c) \bar\gamma_4^2$ (for
$\gamma_6 \gg \gamma_5 = \gamma_4 \gg \gamma_3$),
where $\bar x$ denotes the average value of the quantity $x$ during
the passing process. This estimation is only good for a not very long
passing time, so that the variables do not change much during the
crossing period. This is the case for the rear shell, but not for the
forward shell. For example, for a thin rear shell, $\Delta \sim R
/\gamma_6^2$, and $t_{\Delta_6} \sim (R/c) (\bar\gamma_4 /
\gamma_6)^2$. This is much less than the
dynamical time of the shell propagation in the fixed frame, $R/c$, as
long as $\gamma_4 \ll \gamma_6$. For a thick rear shell, the crossing
time is even shorter since the shell is not spreading.
For the forward shell, however, one has
$t_{\Delta_3}\sim (R/c)$, which is very long. Therefore it is evident
that $t_{\Delta_6}$ is always shorter than $t_{\Delta_3}$, and we use
it to define the end of the collision stage. The radius $R_\Delta$ is
then defined as $\sim R_{col} + ct_{\Delta_6}$, and we denote the
corresponding parameters at this radius with a subscript ``$\Delta$'',
e.g., $e_{\Delta,3}$, $e_{\Delta,4}$, $\gamma_{\Delta,3}$,
$\gamma_{\Delta,4}$, etc. These quantities are recorded from the code
and are inputs to the dynamical problem of the next stage (see below).

\subsubsection{Relaxation stage}

Beyond $R_\Delta$, the reverse shock passes across the rear shell,
and the process enters the dynamical relaxation stage.
After this point, there is no further significant energy input into the
region (4+5).  Since $e_4>e_3$ is still satisfied at $R_\Delta$, the
forward shock B still exists and propagates into the forward shell as
long as the region (4+5) is separated from the region (2+3)
super-sonically. The region (4+5) then expands adiabatically, since
there is practically no heat flow
across the shock. The internal energy density in the region (4+5)
decreases according to the adiabatic law, i.e., $e_4 \propto
U_{4+5}/V' \propto (V')^{-\hat\gamma} \propto R^{-4} \gamma_4^{4/3}$
for $\hat\gamma=4/3$. As a result, $e_4/e_3$ as well as $\gamma_{43}$
drops rapidly towards unity. During this process, $\gamma_3$ increases
and $\gamma_4$ drops, until the whole region (2+3+4+5) reaches the
same velocity, and contacts sub-sonically. The shock (B) dissappears
and the relaxation stage ends, at the radius defined by $R_f$.

During the relaxation stage, the shock conditions (\ref{21}),
(\ref{gam43}) and (\ref{n43}) still hold. Again one needs to quantify
the dynamical evolution of $\gamma_3$ and $\gamma_4$. Unlike the previous
stage, $\gamma_3$ no longer evolves according to the simple $\propto R^{-3/2}$
form. Let us again write down the global energy conservation of the
whole system, i.e., $\gamma_{0,6} M_{0,6}c^2 + \gamma_{0,3} M_{0,3}c^2
+M_2 c^2=\gamma_4 (M_{0,6}c^2+M_4c^2+U_{4+5}) + \gamma_3 (M_{3,0}c^2
-M_4c^2 +M_2c^2+U_{2+3})$.
Noticing that the shock heating energy produced at (A) is only consumed
in the region (2+3), i.e., $U_{2+3} \simeq (\gamma_3-1)M_2c^2$, this gives
\ba
(\gamma_{0,6}-\gamma_4)M_{0,6}-(\gamma_4-\gamma_3)M_4 -\gamma_4
U_{4+5} \nonumber \\
 +(\gamma_{0,3}-\gamma_3) M_{0,3}=(\gamma_3^2-1)M_2~.
\label{stg:4}
\ea
This is even more complicated than (\ref{stg:3}). As the relaxation
goes on, $\gamma_4$ gets closer to $\gamma_3$ and $U_{4+5}$
drops rapidly, thus both the second and the third terms decline.
The first term, which is essentially the injected energy, starts to
influence the blastwave dynamics, just like the Poynting-flux
injection or the mild injection (stage 2) cases (cf. eqs.[\ref{E}] and
[\ref{stg:2}]).

The relaxation is a complicated process. Strictly speaking, one needs to
solve (\ref{stg:4}) and (\ref{gam43}) to find a self-consistent
solution for $\gamma_3$ and $\gamma_4$. Here we perform a simpler but
adequate treatment. Since $\gamma_3$ is boosted up relative to the
non-injection case, we assume that
\be
\gamma_3^2 \propto e_3 \propto R^{-3+\delta}~,
\label{delta}
\ee
where $\delta$ is an unknown positive index. At $R_f$,
the internal energy density of the whole fireball is then $e_{3,f} =
e_{\Delta,3} (R_\Delta/R_f)^{3-\delta}$. Since beyond $R_f$ the fireball
evolves as if the initial total energy is
$E_{0,6}+E_{0,3}=E_{sh,6}+E_{sh,3}$, then $e_{3,f}$ can be also
expressed in an adiabatic expansion form, but as if the energy density
at $R_\Delta$ is a factor $(1+E_{sh,6}/E_{sh,3})$
times larger, i.e., $e_{3,f}=(1+E_{sh,6}/E_{sh,3})e_{\Delta,3}
(R_\Delta/R_f)^3$. This gives
\be
\delta=\frac{\ln\left(1+\frac{E_{sh,6}}{E_{sh,3}}\right)}
{\ln \left(\frac{R_f}{R_\Delta}\right)}~.
\ee
The last task is to determine $R_f$. In the relaxation stage, $e_3$
can be expressed as $=e_{\Delta,3}(R_\Delta/R)^{3-\delta}$, according
to the assumption made. On the other hand, $e_4$ evolves according to the
adiabatic law, i.e., $e_4=e_{\Delta,4} (R_\Delta/R)^4 (\gamma_4 /
\gamma_{\Delta,4})^{4/3}$. One then has $e_4/e_3=(e_{\Delta,4}
/e_{\Delta,3})(R_\Delta/R)^{(9+\delta)/3} (\gamma_{\Delta,3}
/\gamma_{\Delta,4})^{4/3} (\gamma_4/\gamma_3)^{4/3}$. At $R=R_f$,
$e_4/e_3=1$ and $\gamma_4/\gamma_3=1$ by definition. Noticing again
equation (\ref{delta}), the relaxation stage ends at
\be
R_f=R_\Delta \left(\frac{e_{\Delta,4}}{e_{\Delta,3}}\right)^{1/3}
\left(\frac{\gamma_{\Delta,3}}{\gamma_{\Delta,4}}\right)^{4/9}
\left(1+\frac{E_{sh,6}}{E_{sh,3}}\right)^{-1/9}~.
\label{Rf}
\ee
Since the parameters at $R_\Delta$ are in principle known (\S3.3.3),
equations (\ref{delta})-(\ref{Rf}), together with (\ref{21}) and
(\ref{gam43}), adequately describe the dynamics in the relaxation
stage.

Some crude estimates match the numerical results. With $e_4 \propto R^{-4}
\gamma_4^{4/3}$, $e_3 \propto R^{-(3+\delta)}$ and $\gamma_{43}^2
\propto e_4/e_3$ (eq.[\ref{gam43}]), one can get, roughly, $\gamma_4
\propto R^{-6}$, a very steep declination (Fig.4a).
As long as the shock (B) still exists, new electrons are accelerated,
and strong emission from the region heated by the shock (B) continues
throughout the relaxation stage. At the rear end of the shell
(the region 5), since there is no longer a reverse shock (C), no new
electrons are accelerated. However, the already accelerated electrons
are still hot and radiate synchrotron emission. The random Lorentz
factor in the region 5, which we still denote as $\gamma_{_{\rm C}}$
(although the shock (C) no longer exist), also drops rapidly. Since $e_5
= e_4 \propto (V')^{-\hat\gamma} \propto R^{-4}\gamma_4^{4/3}$, and
$n_5 \propto (V')^{-1} \propto R^{-3}\gamma_4$, one has $\gamma_{_{\rm
C}} \propto (V')^{-\hat\gamma+1} \propto R^{-1}\gamma_4^{1/3} \propto
R^{-3}$ (Fig.4a). Since $e_4/e_3$ no longer $\gg 1$, there is no
simple scaling-law estimation of $\gamma_{_{\rm B}}$ at this stage.

\subsubsection{Post-relaxation stage}

After the relaxation ends, one has $\gamma_3=\gamma_4$, the original
$U_{4+5}$ has died out, and the  new thermal energy produced at
the shock (A) is consumed in the whole shell (2+3+4+5), i.e.,
$U_{2+3+4+5} \simeq (\gamma_3 -1)M_2c^2$. The global energy conservation
now reads $\gamma_{0,6} M_{0,6}c^2 + \gamma_{0,3} M_{0,3}c^2+M_2c^2 =
\gamma_3 (M_{0,6}c^2+M_{3,0}c^2+U_{2+3+4+5})$, which is
\be
(\gamma_{0,6}-\gamma_3)M_{0,6} +(\gamma_{0,3}-\gamma_3)
M_{0,3}=(\gamma_3^2-1)M_2~.
\label{stg:5}
\ee
This can be also obtained by removing the second and third terms
in (\ref{stg:4}), as required by the physical relaxation process.
Equation (\ref{stg:5}) is a standard adiabatic expansion equation with
a total energy of $\gamma_{0,6}M_{0,6}c^2+\gamma_{0,3}M_{0,3}c^2= E_{\rm
inj} + E_{\rm imp}$. At this stage, the only shock that accelerates new
electrons is the shock (A), and the emission from it is well described
by the standard afterglow theory.

Although shocks (B) and (C) no longer exist, their already accelerated
electrons are still present and emitting. The random Lorentz
factors in both regions (4 and 5), which we still denote as
$\gamma_{_{\rm B}}$ and $\gamma_{_{\rm C}}$, evolve as $\propto
(V')^{-\hat\gamma+1} \propto R^{-1}\gamma_3^{1/3} \propto
R^{-3/2}$ (Fig.4a).

\begin{figure*}
\centerline{\psfig{file=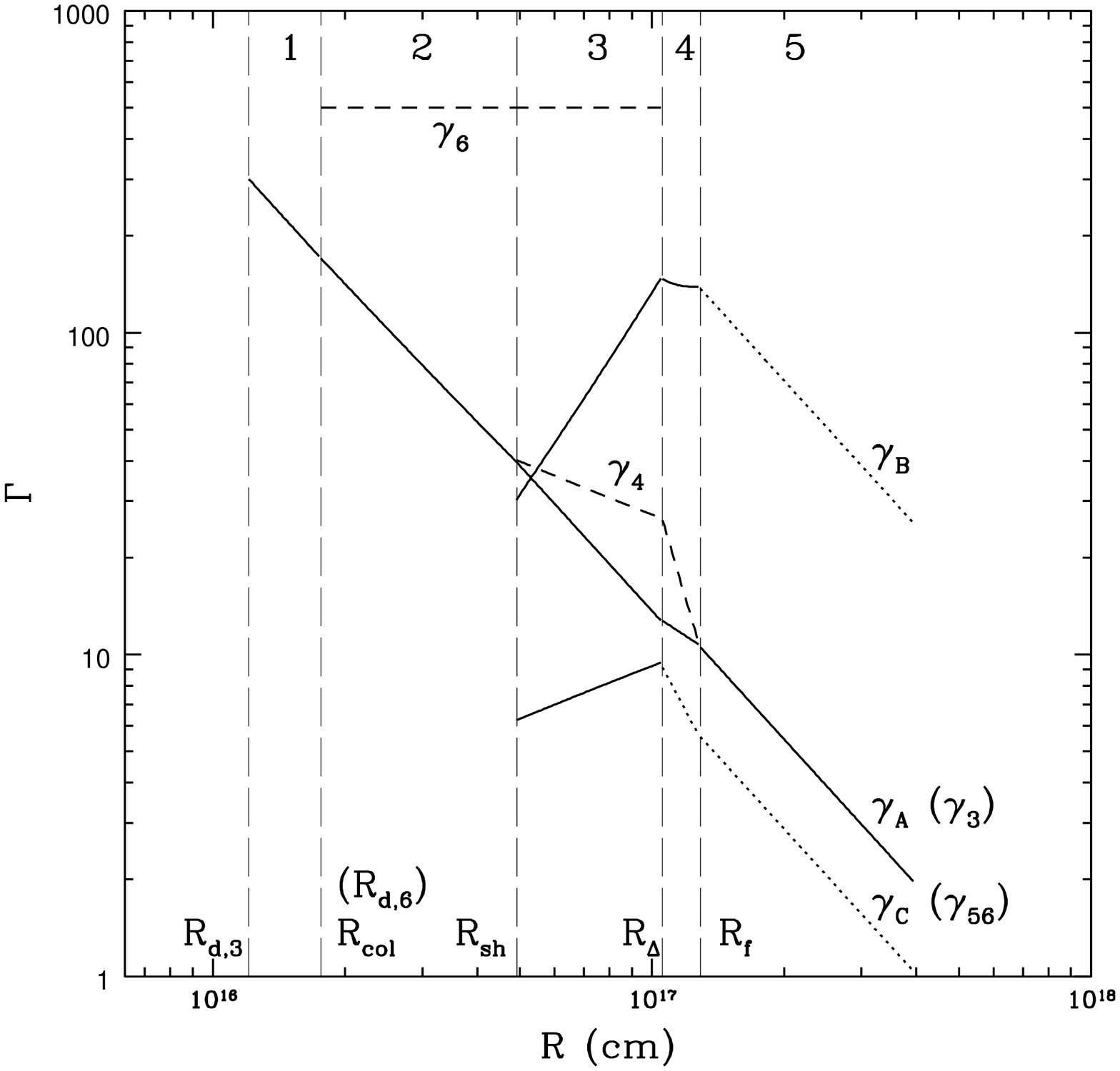,height=8.0cm}
\psfig{file=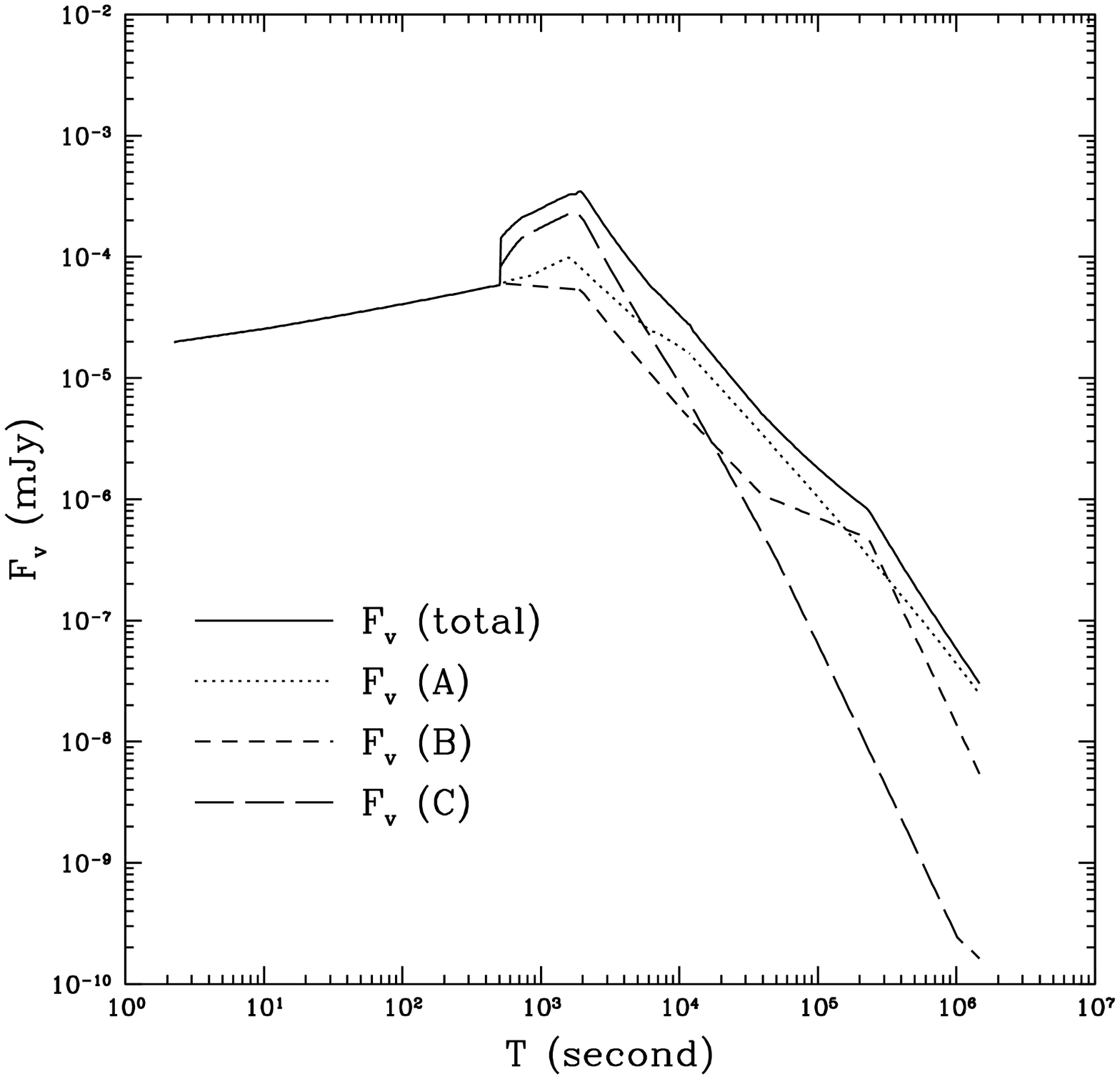,height=8.0cm}}
\caption{An exapmle of the kinetic-energy-dominated shell injection
case. Parameters: for the impulsive shell 3, total energy $E_{0,3}=10^{51}$
ergs, dimensionless entropy $\gamma_{0,3}=300$; for the injective shell 6,
$E_{0,6}=5 \times 10^{50}$ ergs, and $\gamma_{0,6}=500$. The central engine
activity time for the rear shell is $\delta T_6=100$ s, the time interval
between ejecting the two shells is $\Delta T_{36}=10$ s. ISM baryon number
density $n_1=1 ~{\rm cm^{-3}}$, electron spectral index $p=2.5$.
Equipartition parameters $\epsilon_e=0.1$, and $\epsilon_B =0.02$.
The source redshift  $z=1$. (a) Evolution of various Lorentz factors
with blastwave radius $R$. $\gamma_{_{\rm A}}$, $\gamma_{_{\rm B}}$, and
$\gamma_{_{\rm C}}$ are the random Lorentz factors of the emission sites
heated by the shocks (A), (B), and (C), respectively. The curves are solid
when the shocks are on, and are dotted when they are off.
$\gamma_3$, $\gamma_4$, and $\gamma_6$ are the bulk Lorentz factors of the
regions (2+3), (4+5), and 6, respectively (cf. Fig.2). Four critical radii,
i.e., $R_{col}$, $R_{sh}$, $R_{\Delta}$ and $R_f$, separate the process into
5 stages, as marked in the plot. (b) The injection lightcurve in the optical
band ($\nu = 10^{14}$ Hz). The emission components from all the three sites
as well as the total flux are plotted. }
\end{figure*}
\begin{figure*}
\centerline{\psfig{file=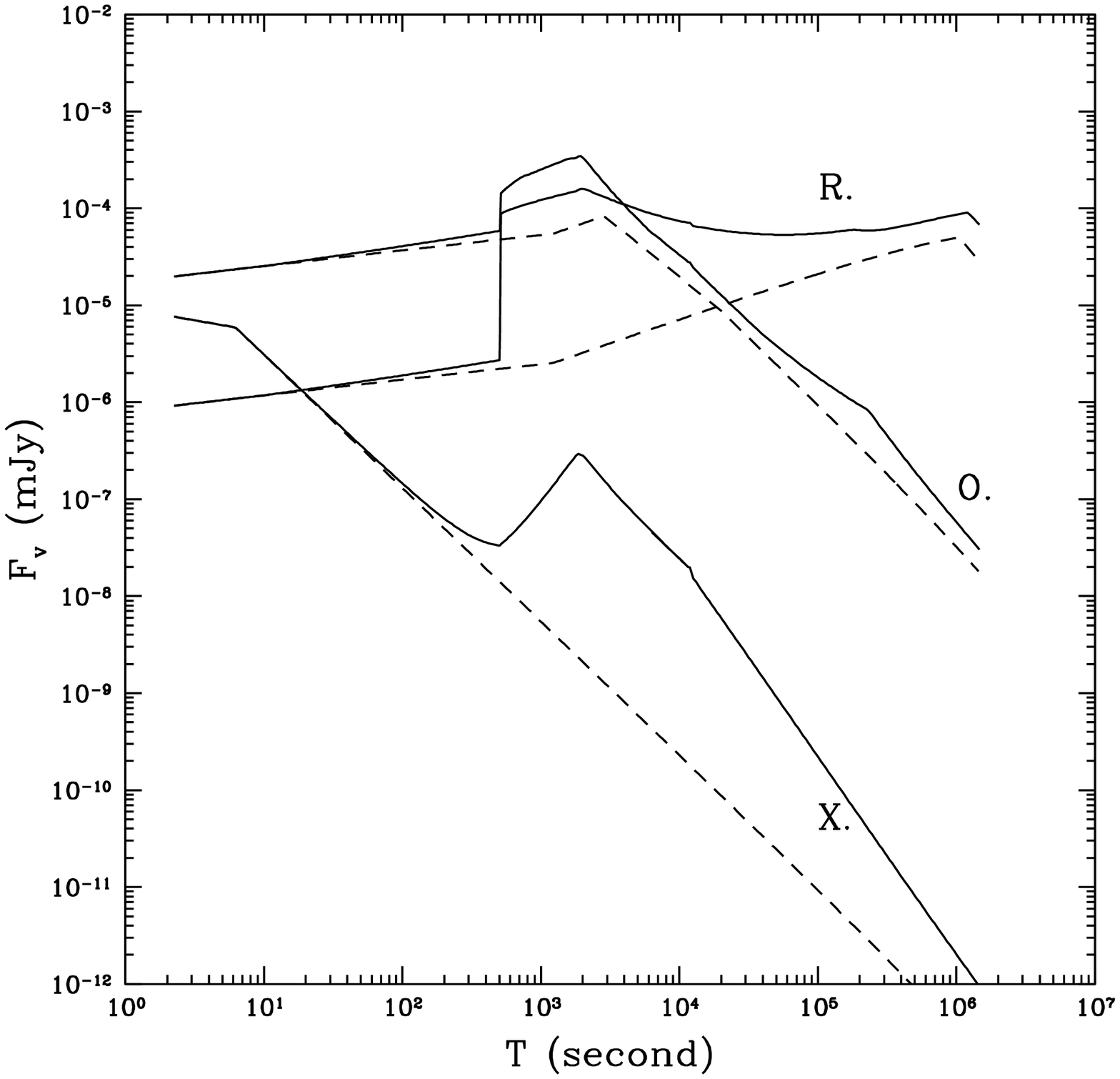,height=8.0cm}
\psfig{file=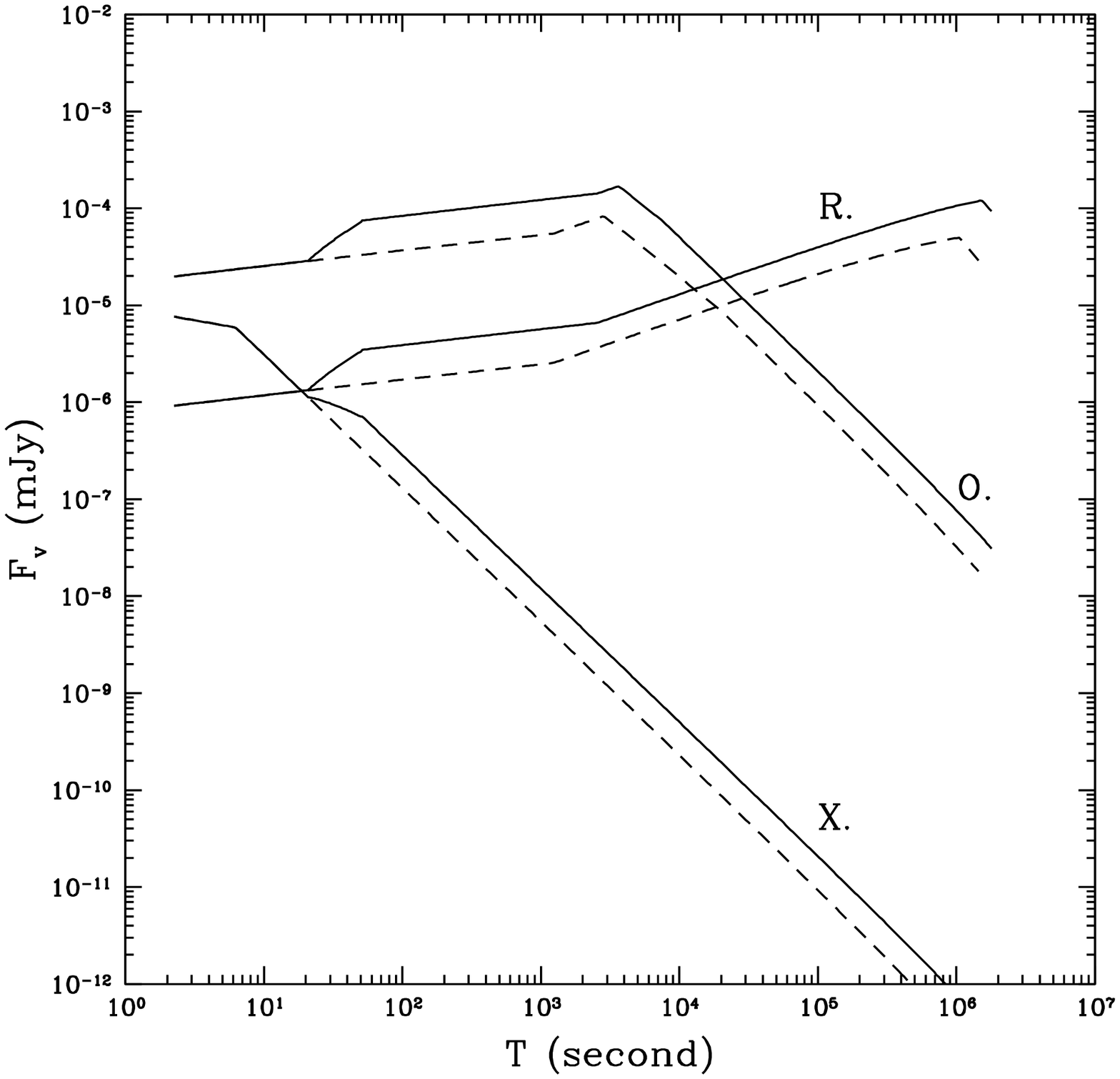,height=8.0cm}}
\caption{Broadband injection signatures (X-ray, optical, and radio)
for the injection of a kinetic-energy-dominated shell. Solid lines
are lightcurves with injection, compared to the dashed line case without
injection. (a) A ``violent'' injection involving formation of new
shocks. All the parameters are the same as adopted in Figure 4.
(b) A mild injection. Injective shell parameters are $E_{0,6}=1
\times 10^{51}$ ergs, $\gamma_{0,6}=200$. The central engine
activity time for the rear shell is $\delta T_6=20$ s, the time interval
between ejecting the two shells is $\Delta T_{36}=20$ s. Other
parameters are the same as in Fig.4 and Fig.5a. Notice that although
the total energy for the shell 6 is higher in the present case, the
relative Lorentz factor between the two shells is much smaller, and
the injection is mild.  }
\end{figure*}

\subsection{Synchrotron radiation and injection lightcurves}

With the shell-merging dynamics quantified above, we can calculate the
injection signatures in the afterglow lightcurves. We have developed a
code to calculate the dynamics of the shell merging process as well as
the emission from various shocks. Since the normal afterglow emission
is generally well described by synchrotron radiation, we have neglected
the high energy spectral components, such as the electron inverse Compton
and the emission from the protons. Although these components could be
important in the MeV to GeV range (see Zhang \& \Mesz~ 2001b for a
detailed discussion), usually synchrotron emission is adequate to describe
the low energy afterglow (X-ray, optical and radio), at least in the early
afterglow phase when injection may be expected. Nonetheless, the
electron cooling due to the synchrotron self-Compton is incorporated
through the Y parameter (see, e.g. Panaitescu \& Kumar 2000; Sari \&
Esin 2001; Zhang \& \Mesz~ 2001b).

The random thermal energy in a certain region is an essential
parameter in calculating the synchrotron spectrum and the lightcurves.
In stages 1 and 2, the only emitting site is the region heated by the
shock (A). Starting from the stage 3, the emission sites increase to a
total number of three, corresponding to the three shocks (although the
shocks (B) and (C) die out in the stages 5 and 4, respectively). When
the shocks are on, the random Lorentz factors at the shocks (A) and (C)
are essentially the Lorentz factor of the shocks, i.e., $\gamma_{_{\rm
A}}=\gamma_3$, and $\gamma_{_{\rm C}} = \gamma_{56}$.
For the shock (B), since the unshocked region (region 3)
is also hot, the random Lorentz factor is not simply $\gamma_{43}$,
but is rather $\gamma_{_{\rm B}}=e_4/n_4m_pc^2$. When the shocks (B)
and (C) die out, the random Lorentz factors in the relevant regions
are still defined as $\gamma_{_{\rm B}}=e_4/n_4 m_pc^2$ and
$\gamma_{_{\rm C}}=e_5/n_5 m_pc^2$, which evolve
adiabatically as discussed in \S3.3.4 and \S3.3.5. To calculate the
synchrotron emission, we assume uniform values of the equipartition
factors, i.e., $\epsilon_B$ and $\epsilon_e$, as well as the same
electron power law index $p$ in the three shock-heated regions.
This is a crude approximation, but it is the simplest case which
allows a straightforward comparison of the relative importance
of the three shocks. The magnetic field in the region $i$
is then $B_i=(8\pi\epsilon_B e_i)^{1/2}$. Assuming that the injection
factor of the electrons is unity, the characteristic injected
electron Lorentz factor is then $\gamma_{m,{\rm X}}=\epsilon_e
[(p-2)/(p-1)] (m_p/m_e) \gamma_{_{\rm X}}$, where $\gamma_{_{\rm X}}$
is the random Lorentz factor at the region X (A, B or C), $m_p/m_e$ is
the proton-electron mass ratio. Cooling frequencies are determined
from the cooling electron energy, which is defined by the equality
between the comoving time and the electron radiative cooling time.
For the regions where the shocks are
on and keep accelerating electrons, the synchrotron emission
spectrum is calculated according to the four-segment broken power law
description (Sari, Piran \& Narayan 1998). For those regions where no
new electrons are accelerated (region 5 in the stages 4 and 5, and
region 4 in the stage 5), the synchrotron spectrum is modified by a
sharp cut-off above the cooling frequency (although strictly speaking
the cut-off should be exponential).

One crucial point in calculating the lightcurves is that the radiation
from the three shocks which is emitted at the same fixed-frame time will
reach the observer at different observer times. The main reason is that
the region (4+5) usually moves faster (with $\gamma_4$) than the region
(2+3) (with $\gamma_3$), so that for a same time interval in the fixed
frame, $d t$, the emission from the sites heated by the shocks (B) and (C)
reaches the observer (apart from the cosmological time dilation
effect) in an interval of $\sim dt/\gamma_4^2$, while the
emission from the shock (A) reaches the observer in an interval of
$\sim dt/\gamma_3^2$. In our calculations, this effect is taken into
account approximately. Defining the observer's time when the front of
shell 3 reaches $R_{sh}$ to be $T_{sh}$, we can approximate the
observer's times corresponding to the regions (4+5) and (2+3) after
$T_{sh}$ to be roughly $T_4 \sim T_{sh}+ (R-R_{sh})/4\gamma_4^2c$,
and $T_3 \sim T_{sh}+ (R-R_{sh})/4\gamma_3^2c$,
respectively\footnote{Strictly speaking, since the region (4+5) trails
behind the region (2+3), the time delay due to the emission from
the region (4+5) crossing the shell 3 ought to be taken into
account. Since the shell width $\Delta_3 \ll R$, this effect may be
neglected as long as $(R-R_{sh}) \gg \Delta_3$.}. We see that the two
times are different in the stages 3 and 4, but are the same in stages
1, 2 and 5. The overall effect of this correction is to ``squeeze'' the
lightcurve signal from the fast moving region (4+5) during stages 3 and 4.

An example of a shell collision is presented in Figure 4. In
Fig.4a, we plot the Lorentz factor evolution of various shocks and
regions with respect to the blastwave radius, or equivalently the
fixed-frame time $t\sim R/c$. This allows a clear assessment of the five
stages analyzed in \S3.3. Figure 4b presents an indicative optical
lightcurve, identifying the contributions from the three emitting sites.
The final lightcurve is complicated, with a few features which are caused
by different emission sites. To make more sense of this numerical lightcurve,
it is helpful to perform some further analysis. The emission component
from the site related to the shock (A) is a standard ``low frequency''
lightcurve (Sari et al. 1998), except for a gentle flattenning around
$10^4$ s, which corresponds to the relaxation boosting-up of
$\gamma_3$ in the stage 4 (Fig.4a). The emission components
corresponding to the shocks (B) and (C) are more complicated. The
lightcurve slopes in various sections depend on the (observer's) time
dependence of the characteristic frequency, $\nu_m \propto \gamma_i
\gamma_{_{\rm X}}^2 B'$ for slow-cooling (which is usually the case) and
$\nu_c$ for fast-cooling, and the peak spectral flux, $F_{\nu,m}
\propto n_i \gamma_i B' R^3$. Here $\gamma_i$ denotes the bulk
Lorentz factors ($\gamma_3$ or $\gamma_4$) and $\gamma_{_{\rm X}}$ denotes
the random Lorentz factors ($\gamma_{_{\rm A}}$, $\gamma_{_{\rm B}}$,
and $\gamma_{_{\rm C}}$). Any change of the scaling law (with respect
to $T$) of any parameter (e.g., $\gamma_{_{\rm X}}$, $\gamma_i$, $n_i$,
$B'$, $R$) results in a break in the lightcurves, and the lightcurve
slope also depends on the spectral segment in which the observing frequency
lies. In the specific example presented in Fig.4b, the climbing-up
segment in the lightcurve (C) and the flat segment in the lightcurve
(B) correspond to the violent injection stage 3, but the timescale
is squeezed due to the increase of $\gamma_4$ with respect to
$\gamma_3$. The later drop-offs in both lightcurves correspond to the
relaxation stage. Finally, all three lightcurves enter the
post-relaxation stage at a same observer's time. There could be more
breaks within the same stage. For example, the break on the lightcurve
(B) in the post-relaxation stage is due to the transition from the (1/3)
spectral segment to $[-(p-1)/2]$ spectral segment. Two remarks can be
made here. First, although in the stage 5 all the three random
Lorentz factors decay as $\propto R^{-3/2}$, the lightcurve slopes on
(B) and (C) in the $[-(p-1)/2]$ spectral segment are steeper than the
slope on (A) in the same spectral segment. This arises from the different
scaling law in $F_{\nu,m}$, more particularly in $n$.
For adiabatic expansion in the regions (B) and (C),
$n_i\propto (V')^{-1} \propto R^{-3}\gamma_i \propto R^{-9/2}
\propto T^{-9/8}$. In the region (A), $n_2 \propto \gamma_3 \propto
R^{-3/2} \propto T^{-3/8}$. The lightcurve slopes for (B) and (C)
then steepen by 0.75 relative to that of (A)\footnote{In Fig.4b, the
steepening is slightly smaller than 0.75. This is because the lightcurve
(A) is in the $(-p/2)$ spectral regime, which has steepened by itself.}.
Also the whole lightcurves (B) and (C) during the stage 5 also tilt by
0.75, which explains why the crossing of $\nu_m$ in the lightcurve (B)
looks a little unfamiliar.
Second, although the random Lorentz factor in the site (B) is
much higher than those in sites (A) and (C), the flux level is not,
mainly because the density $n_4$ is smaller than $n_2$ and $n_5$
during the violent collision process.

From Fig.4b, one can see that the injection signatures of a violent
collision are rather complicated. There could be more than one signature
which may be related to different emission sites. This gives rise to a
whole set of injection patterns. For example, the sharp decline of the
site (C) emission in the relaxation stage, given an appropriate set
of parameters, could be an explanation for the ROTSE prompt optical flash
observation in GRB 990123, which is conventionally interpreted as the
cooling from the reverse-shock-heated region in an impulsive shell
case (\Mesz~ \& Rees 1997a, 1999; Sari \& Piran 1999, Kobayashi 2000).
The modest sticking-out of the optical lightcurve (B) around several
$10^5$ s (Fig.4a) seems to have been observed in some GRB lightcurves,
although the chromatic nature of this rise rules it out as a possible
explanation of the achromatic bump observed in GRB 000301C, which can
be interpreted as a gravitational micro-lensing event (Garnavich et
al. 2000).

The main characteristic which can be used to differentiate a
post-injection model from other possible mechanisms leading to bumps
in the light curve is the broad-band behavior. Since the injection
processes produce global dynamics changes, they will leave imprints at
all wavebands. In Figure 5 we present the broad-band injection
lightcurves for (a) a violent injection; (b) a mild injection
throughout. We can see that the mild injection case is similar to
the Poynting-flux-dominated injection case (Fig.1), as has been
discussed in \S3.3.2. The violent injection case shows rather
different and complex behaviors. The rising times in various bands are 
essentially the same, but the bump shapes are quite different, mainly
because different shocks contribute differently in different bands.

\section{Summary \& Discussion}
\label{sec:sum}

In this paper, we have analyzed the observational consequences of a
GRB fireball which receives an additional energy injection from the
central engine during the afterglow phase. We have discussed the cases
where the injection is either a continuous Poynting-flux-dominated
wind, or a kinetic-energy-dominated matter shell. For the latter, we
discuss the cases when the collision is either ``mild'' or
``violent'', distinguished by whether additional shocks persist and
additional emission sites arise in the outflow. We reach the following
general conclusions:

1. A post-burst injection energy $E_{\rm inj} \sim E_{\rm imp}$ is
generally required to cause a noticeable signature in the afterglow
lightcurves. This is especially so for the Poynting flux injection
case and for the mild kinetic energy injection case. For a violent
collision, a faster rear shell (i.e. $\gamma_6 \gg \gamma_3$) can
compensate for a lower rear shell total energy (Fig.3).

2. For the kinetic-energy injection case, we have explicitly studied
the three-shell interaction process, described by a set of equations
(eqs.[\ref{21}]-[\ref{56}]). We find that the formation of an
additional pair of shocks from the collision of two shells is not as
common as usually thought. Depending on the status of the rear shell
when the collision occurs (spreading or not), a minimum relative speed
between the two colliding shells is required, as shown in Figure 3. In
many cases, such as expected in the internal shock model where slow
trailing shells catch up with the fast shells when the latter are
decelerated, the relative speed between the shells are usually not
high enough, and the injections are likely to be mild. We expect that
a violent injection most likely arises from a late injection of a high
entropy shell from the central engine. Since the baryon loading at
late times is expected to be smaller than that at earlier times, this
is not unusual given a long-lived erratic central engine.

3. If no additional emission sites form during the injection, such as
the case of the Poynting-flux-dominated flow, the injection signature
is usally very gradual (Fig.1 and Fig.5b). In the case of a violent
injection, the signatures are complicated and some could be very
abrupt.  In either case, the final Lorentz factor of the blastwave, as
well as the emission flux level, are boosted up with respect to
the case without post-injection.

4. A prominent characteristic of the post-injection process is the
conspiratorial variation of the flux in all bands (Fig.1
and Fig.5), due to the changies in the dynamics of the emitting
region.  This provides a strong criterion for judging whether a bump
signature is due to an injection or to other reasons. For example, the
supernova component (Bloom et al. 1999; Reichart 1999; Galama et
al. 2000) and the dust echo (Esin \& Blandford 2000) interpretations
do not predict prominent changes in X-ray and radio lightcurves.
The gravitational micro-lensing model may be more difficult to
differentiate from a post-injection event, since it is essentially an
achromatic feature, although it can have some chromatic features
(Granot \& Loeb 2001). The Poynting-flow injection case is also
achromatic. The differentiation between the two possibilities can be
only based on the shape of the bump. For example, the micro-lensing
model predicts a more abrupt (shorter duration) feature than does the
Poynting-injection. Another feature of micro-lensing is that the afterglow
lightcurve will resume the level of no magnification when the lensing
event ends, while in the injection case, the global energetics is
always boosted up. Prospects of differentiation look better for a
violent matter-injection case, in which the bumps in different bands
have quite different shapes, although the rising time is essentially
the same (Fig.5a). Overall, future broadband afterglow data with very
good temporal coverages plus a more detailed modelling can in
principle differentiate the injection signature from the other
possibilities. Within the injection itself, different injection
signatures may provide direct information about the nature of the
injection, and possibly, also about the nature of the central engine.

The treatment presented in this paper is generic and applicable also
to more complicated injection processes. For example, for
multi-shell-collision cases, one simply needs to regard the emitting
gas after each injection as a new ``impulsive'' component, and a newly
injected shell or wind can be treated in the same way. Since
collisions in the internal shock scenario also involve a similar
process (but without the interaction with the ISM), the treatment
presented here may be also usable in studying GRB gamma-ray
lightcurves in the prompt phase.  The shell merging is a very
complicated process, and here we have used some approximations leading
to a semi-analytic description of the whole process. For example, we
have assumed that the decelerating shell has a uniform density and
speed, while a more accurate treatment ought to take into account the
Blandford-McKee (1976) self-similar profiles. This approximation
nonetheless provides a first-order treatment to the problem. Another
assumption we have made is that the total internal energy in each
region is solely due to the shock heating. Strictly speaking,
adiabatic loss due to expansion should be taken into account
(e.g. Panaitescu et al. 1998).  However, as long as shock heating is
going on, the adiabatic loss is not important. Only when shock heating
ceases (e.g., region 5 in the stages 4 and 5, and region 4 in the
stage 5), do we consider the adiabatic loss. Furthermore, in all our
calculations, the effects concerning spreading in the photon arrival
time due to the curvature of the shells as well as their non-zero
thicknesses have been ignored. These effects are essential in
more detailed afterglow modeling (e.g. Granot, Piran \& Sari 1999),
which tend to smear out features on the lightcurves. With these
effects taken into account, we expect that the sharp features shown in
Fig.5a ought to be smeared out.

Finally we compare our results with the previous relevant treatments.
Rees \& \Mesz~ (1998), and Panaitescu et al. (1998) have discussed the
``refreshed'' injection scenario in which all the mini-shells are
assumed to be ejected simultaneously, but with a power-law
distribution of $\Gamma$. Although such a scenario is of interest and
can be modeled relatively easily, it is unclear whether it is a closer
representation of reality.  The Poynting-flux-dominated injection
within the pulsar central engine model has been discussed earlier by
Dai \& Lu (1998), but their indicative lightcurve was too abrupt.
Their recent treatments
(Dai \& Lu 2001; Wang \& Dai 2001) generally agree with the gradual
evolution picture obtained in this paper. For the shell collision
injection case, the most relevant precedent discussion is the one by
Kumar \& Piran (2000). Although our results generally agree with
theirs on, e.g., the hydrodynamic treatments, our shock forming
condition (Fig.3) is more demanding than theirs. In fact, for the
specific example they presented, i.e., $E_{0,6}\sim E_{0,3}$, and
$\gamma_{36} \sim 1.25$, the injection is only mild according to our
criterion, while they argue that two shocks will form and propagate
into the two shells. This discrepancy may be due to their not having
included the hydrodynamics of the three-shell interaction. They also
presented an indicative injection lightcurve (their Fig.5). However,
this lightcurve is more analogous to our mild-injection signature
(Fig.5b). As discussed in \S3.4, a violent injection tends to generate
a more complicated lightcurve with possibly more than one signature, as
indicated in Fig.4b and Fig.5a.

\acknowledgments{We are grateful to Pawan Kumar, Alin Panaitescu and
Re'em Sari for informative correspondence and Oleg Kargaltsev for
discussions, to the referee for a detailed and helpful report,
and to NASA NAG5-9192 and NAG5-9153 for support.}


\begin{references}
\reference{} Blackman, E. G., \& Yi, I. 1998, ApJ, 498, L31
\reference{} Blandford, R, \& McKee, C. 1976, Phys. Fluids, 19, 1130
\reference{} Bloom, J. S. et al. 1999, Nature, 401, 453
\reference{} Chiang, J., \& Dermer, C. D. 1999, ApJ, 512, 699
\reference{} Dai, Z. G., \& Lu, T. 1998, Phys. Rev. Lett., 81, 4301
\reference{} -----. 2001, A\&A, 367, 501
\reference{} Esin, A. A., \& Blandford, R. 2000, ApJ, 534, L151
\reference{} Galama, T. et al. 2000, ApJ, 536, 185
\reference{} Garnavich, P, Loeb, A, \& Stanek, K. 2000, ApJ, 544, L11
\reference{} Granot, J., \& Loeb, A. 2001, ApJ, 551, L63
\reference{} Granot, J., Piran, T., \& Sari, R. 1999, ApJ, 513, 679
\reference{} Huang, Y. F., Dai, Z. G., \& Lu, T. 1999, MNRAS, 309, 513
\reference{} Kennel, C. F., \& Coroniti, F. V. 1984, ApJ, 283, 694
\reference{} Kobayashi, S. 2000, ApJ, 545, 807
\reference{} Kobayashi, S., Piran, T. \& Sari, R. 1999, ApJ, 513, 669
\reference{} Kumar, P., \& Piran, T. 2000, ApJ, 532, 286
\reference{} Lyutikov, M., \& Blackman, E. G. 2001, MNRAS, 321, 177
\reference{} \Mesz, P., Laguna, P. \& Rees, M.J. 1993, ApJ, 415, 181
\reference{} \Mesz, P., \& Rees, M. J. 1997a, ApJ, 476, 232
\reference{} -----. 1997b, ApJ, 482, L29
\reference{} -----. 1999, MNRAS, 306, L39
\reference{} Panaitescu, A., \& Kumar, P. 2000, ApJ, 543, 66
\reference{} Panaitescu, A., \Mesz, P., \& Rees, M. J. 1998, ApJ, 503,
             314
\reference{} Piran, T. 1999, Phys. Rep., 314, 575
\reference{} Piran, T., Shemi, A. \& Narayan, R. 1993, MNRAS, 263, 861
\reference{} Rees, M. J. \& \Mesz, P. 1998, ApJ, 496, L1
\reference{} Reichart, D.E. 1999, ApJ, 521, L111
\reference{} Sari, R., \& Esin, A. A. 2001, ApJ, 548, 787
\reference{} Sari, R., \& Piran, T. 1995, ApJ, 455, L143
\reference{} -----. 1999, ApJ, 517, L109
\reference{} Sari, R., Piran, T., \& Narayan, R. 1998, ApJ, 497, L17
\reference{} Spruit, H. C., Daigne, F., \& Drenkhahn, G. 2001, A\&A,
             369, 694
\reference{} Usov, V. V. 1994, MNRAS, 267, 1035
\reference{} -----. 1999, in ASP Conf. Ser. 190. Gamma-Ray Bursts: The
             First Three Minutes, ed. J. Poutanen \& R. Svensson (San
             Francisco: ASP), 153
\reference{} Wang, W. \& Dai, Z.G. 2001, Chin. Phys. Lett., 18, 1153
\reference{} Zhang, B., \& \Mesz, P. 2001a, ApJ, 552, L35
\reference{} -----. 2001b, ApJ, 559, 110

\end{references}
\end{document}